\pdfoutput=1
\documentclass[preprintnumbers,floatfix,aps]{revtex4}
\usepackage{amsmath}
\usepackage{amsfonts}
\usepackage{amssymb}
\usepackage{graphicx}
\usepackage{epsfig}
\usepackage[hang,nooneline]{subfigure}
\usepackage{color}

\newcommand{\sech}{{\rm sech}}

\newcounter{fig}

\begin{document}

\title{$\mathcal{PT}$ Meets Supersymmetry and Nonlinearity: An Analytically Tractable Case Example}
\author{P. G. Kevrekidis \thanks{%
Email: kevrekid@math.umass.edu}}
\affiliation{Department of Mathematics and Statistics, University of Massachusetts,
Amherst, MA 01003-4515, USA}

\affiliation{Center for Nonlinear Studies and Theoretical Division, Los Alamos National Laboratory, Los Alamos, New Mexico 87545, USA}

\author{Jes\'us Cuevas--Maraver}
\affiliation{Grupo de F\'{\i}sica No Lineal, Departamento de F\'{\i}sica Aplicada I, Universidad de Sevilla. Escuela Polit{\'e}cnica Superior, C/ Virgen de \'Africa, 7, 41011-Sevilla, Spain\\
Instituto de Matem\'aticas de la Universidad de Sevilla (IMUS). Edificio Celestino Mutis. Avda. Reina Mercedes s/n, 41012-Sevilla, Spain}

\author{Avadh Saxena}
\affiliation{Center for Nonlinear Studies and Theoretical Division, Los Alamos National Laboratory, Los Alamos, New Mexico 87545, USA}

\author{Fred Cooper}
\affiliation{Santa Fe Institute, Santa Fe, NM 87501, USA}
\affiliation{Center for Nonlinear Studies and Theoretical Division, Los Alamos National Laboratory, Los Alamos, New Mexico 87545, USA}

\author{Avinash Khare}
\affiliation{Physics Department, Savitribai Phule Pune University, Pune 411007, India}

\date{\today}

\begin{abstract}
In the present work, we combine the notion of $\mathcal{PT}$-symmetry
with that of super-symmetry (SUSY) for a prototypical case example with a complex potential
that is related by SUSY to the so-called P{\"o}schl-Teller potential which is
real. Not only are we able to identify
and numerically confirm the eigenvalues of the relevant problem,
but we also show that the corresponding nonlinear problem,
in the presence of an arbitrary power law nonlinearity, has an exact bright
soliton solution that can be analytically identified and has intriguing
stability properties, such as an oscillatory instability, which
the corresponding solution
of the regular nonlinear Schr{\"o}dinger equation with arbitrary power law nonlinearity
does not possess. The spectral properties and dynamical implications
of this instability are examined. We believe that these findings may
pave the way towards initiating a fruitful interplay between
the notions of  $\mathcal{PT}$-symmetry, super-symmetric partner potentials
and nonlinear interactions.
\end{abstract}

\maketitle

\section{Introduction}

In the past 15 years, there has been a tremendous growth in the number
of studies of open systems bearing both gain and loss, motivated
to a considerable degree by the study of the specially balanced
$\mathcal{PT}$-symmetric dynamical
models~\cite{Bender_review,special-issues,review}.
The original proposal of Bender and collaborators towards the
study of such systems was made as an alternative to the
postulate of Hermiticity in quantum mechanics. Yet, in the
next decade, proposals aimed at the experimental realization
of such $\mathcal{PT}$-symmetric systems found a natural ``home''
in the realm of optics~\cite{Muga,PT_periodic}. Within the latter, the above
theoretical proposal (due to the formal similarity of the Maxwell
equations in the paraxial approximation and the Schr{\"o}dinger
equation) quickly led to a series of experiments~\cite{experiment}.
In turn, these efforts motivated experiments in numerous other areas,
which span, among others, the examination of
$\mathcal{PT}$-symmetric electronic
circuits~\cite{tsampikos_recent,tsampikos_review}, mechanical
systems~\cite{pt_mech} and whispering-gallery microcavities~\cite{pt_whisper}.

In the same spirit, another important idea that has originally
been proposed in a different setting (namely that of
high-energy physics~\cite{avinashfred}) but has recently found
intriguing applications in the context of wave guiding and manipulation in
the realm of optics is that of super-symmetry (SUSY)~\cite{heinr1}.
The main idea is that from a potential with desired
properties, one can obtain a SUSY-partner potential that will be isospectral to
(i.e., possess the same spectrum as) the original one, with the possible
exception of one eigenvalue. In fact, taking the idea one step
further, starting from a desired ground state eigenfunction,
one can design the relevant super-symmetric partner potentials
in a systematic fashion, as discussed, e.g., in~\cite{heinr1},
both for continuum and even for discrete problems. In fact,
more recently, the two ideas (of $\mathcal{PT}$-symmetry, or
anyway non-hermiticity, and
SUSY) have been combined to construct SUSY-partner complex optical
potentials designed to have real spectra~\cite{heinr2}. An expected
application of these ideas that has also started to
be explored (extending the spirit of corresponding
studies in the $\mathcal{PT}$-symmetric setting~\cite{tsamp}) is
in using SUSY transformations to achieve transparent and
one-way reflectionless complex optical potentials~\cite{midya}.

The above works have essentially constrained the interplay of
$\mathcal{PT}$ symmetry and SUSY at the level of linear potentials.
Naturally, however, except for very low optical intensity,
the crystals considered in the relevant applications bear nonlinear
features, e.g., due to the Kerr effect. Hence, our focus in the
present work will be to extend these linear ideas of
$\mathcal{PT}$ symmetry and SUSY to a nonlinear case example. Moreover,
we will select an example that blends two additional characteristics.
On the one hand, one of our super-symmetric partners will constitute
a famous and well-known solvable model in quantum mechanics,
namely the celebrated P{\"o}schl-Teller potential~\cite{poschl,LL}.
On the other hand, it will turn out to be the case that not only the
linear but also the nonlinear variant of the problem will be analytically
solvable, in fact for arbitrary powers of the nonlinearity,
in a special limit and will naturally connect with the linear
solutions of the potential. In what follows, in Sec. II  we will first present
the general theory of linear $\mathcal{PT}$-supersymmetric potentials. Then,
in Sec. III we will consider the special nonlinear solutions and their
asymptotic linear limit reduction. Numerical results will corroborate
the above analytical findings and we will also explore the spectral and
dynamical stability of the nonlinear waveforms. Finally, in section IV,
we summarize our findings and present our conclusions.

\section{A Linear Non-Hermitian Supersymmetric Model}

As is done generally in the theory of SUSY, we consider an
operator $\mathcal{A}$ such that
\begin{eqnarray}
\mathcal{A}=\frac{d}{dx} + W ,
\label{eqn1}
\end{eqnarray}
where $W$ is the super-potential and an operator $\mathcal{B}$ of the form:
\begin{eqnarray}
\mathcal{B}=-\frac{d}{dx} + W .
\label{eqn2}
\end{eqnarray}
It is important to accentuate here (see also~\cite{heinr2}) that
in the case of a complex super-potential $W$, contrary to the
Hermitian case of a real $W$, $\mathcal{B}$ is not a Hermitian adjoint operator
of $\mathcal{A}$ (hence the different symbol). Then, defining the
potentials $V^{\pm}=W^2 \mp W' + E$, with $V^{(+)}=V^{(-)} -2 W'$,
we have that the operators
\begin{eqnarray}
H^{(\pm)} = -\frac{d^2}{d x^2} + V^{(\pm)} - E
\label{eqn3}
\end{eqnarray}
are isospectral (with the exception of the fundamental
mode in the potential $V^{(+)}$ which lacks a counterpart
in $V^{(-)}$. More specifically, the eigenvalues satisfy
$E_n^{(+)}=E_{n-1}^{(-)}$ for $n \geq 1$ (cf. also~\cite{heinr2}).
We note in passing that the eigenvectors of the two cases
are also related, i.e., $u_n^{(-)}=\mathcal{A} u_{n+1}^{(+)}$
and $u_{n+1}^{(+)}=\mathcal{B} u_n^{(-)}$.

Now, assuming that $W=f+ i g$, $V^{(+)}=V_R^{(+)}+iV_I^{(+)}$, $V^{(-)}=V_R^{(-)}+iV_I^{(-)}$ and that $E \in \mathbb{R}$, we find that (cfr.~\cite{bagchi}) the potentials have to satisfy the following conditions:
\begin{eqnarray}
V_R^{(+)} &=& f^2 - g^2 - f' + E ,
\label{eqn4}
\\
 V_I^{(+)} &=& 2 f g  - g' ,
\label{eqn5}
\\
V_R^{(-)} &=& f^2 - g^2 + f' + E ,
\label{eqn4a}
\\
 V_I^{(-)} &=& 2 f g  + g' .
\label{eqn5a}
\end{eqnarray}
The remarkable finding of the linear spectral analysis of~\cite{bagchi}
was that these authors, motivated by the sl$(2,C)$ potential algebra
were able to derive a number of special case examples of simple
functional forms of complex $W$'s which give rise to complex SUSY
potentials. Perhaps the most remarkable of their examples concerns
the super-potential
\begin{eqnarray}
W(x)=\left(m-\frac{1}{2}\right) \tanh(x-c) - i b_I {\rm sech}(x-c),
\label{eqn6}
\end{eqnarray}
which gives rise (assuming hereafter without loss of generality
that $c=0$) to the super-symmetric partners of the form:
\begin{eqnarray}
V^{(+)} &=& \left(-b_I^2-m^2+\frac{1}{4}\right) {\rm sech}^2(x) -2 i m b_I {\rm sech}(x)
\tanh(x),
\label{eqn7}
\\
V^{(-)} &=& \left(-b_I^2 - (m-1)^2 +\frac{1}{4}\right) {\rm sech}^2(x)
-2 i (m-1) b_I {\rm sech}(x)
\tanh(x).
\label{eqn8}
\end{eqnarray}

We chiefly focus hereafter on the remarkable special case of
$m=1$, previously considered, e.g., in~\cite{bagchi2}.
The exceptional characteristic of this case is that
it stems from a real potential $V^{(-)}$ which is well-known
to be exactly solvable in the realm of elementary quantum
mechanics, namely the P{\"o}schl-Teller potential~\cite{poschl,LL}.
While its eigenfunctions can also be written down in an
explicit form by means of hypergeometric functions, here
we will restrict our considerations to the relevant
(bound state) eigenvalues which in the context of the above example
assume an extremely simple form as:
\begin{eqnarray}
E_n^{(-)}=-\frac{1}{4} \left[2 b_I - 2 n - 1 \right]^2.
\label{eqn9}
\end{eqnarray}
Such bound state eigenvalues only exist when $n < b_I - 1/2$.
This, in turn, suggests that for the $+$ superscript potential,
it will be: $E_n^{(+)}=E_{n-1}^{(-)}$, i.e., all the relevant bound
state eigenvalues should also emerge in the $\mathcal{PT}$-symmetric
spectrum of the potential $V^{(+)}$, just as they appear in
the Hermitian (real) spectrum of the potential $V^{(-)}$.
The only eigenvalue that will not be captured by this
relation is $E=-1/4$; see the relevant details on the
spectrum of $V^{(+)}$ below. Furthermore, we expect that, when
varying $b_I$, bound state eigenvalues will emerge as
$b_I$ crosses $0.5$, $1.5$, $2.5, \dots$ in both the
spectra of $V^{(\pm)}$.

\begin{figure}
\begin{center}
\includegraphics[width=6cm]{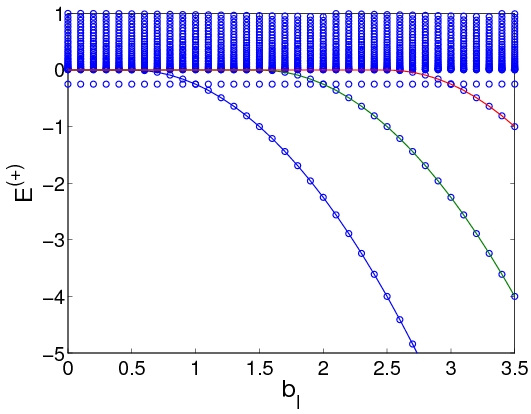}
\end{center}
\caption{The blue circles represent the eigenvalues $E^{(+)}$
of the operator $H^{(+)}$ of Eq.~(\ref{eqn3}), under the $\mathcal{PT}$-symmetric
potential of Eq.~(\ref{eqn7}). The solid lines represent the analytical
predictions based on the potential's super-symmetric partner $V^{(-)}$
corresponding to an analytically tractable P{\"o}schl-Teller potential.}
\label{fig:profile}
\end{figure}

All of these conclusions are fully corroborated by the
results of Fig.~\ref{fig:profile}. The spectrum of
$H^{(+)}$ considered therein turns out to be real,
as may be anticipated by the $\mathcal{PT}$-symmetry of the model,
but more importantly, it turns out to be identical to that
of its super-symmetric P{\"o}schl-Teller partner, as can
be seen from the theoretical lines confirming the bifurcation
of the point spectrum eigenvalues at the locations theoretically
predicted. Finally, indeed, the only eigenvalue that is not
captured is $E=-1/4$ which turns out to be {\it invariant}, under
variations of $b_I$. As a final note, we point
out that generalizations of this potential with arbitrary coefficients
in both the real and the imaginary part were considered in~\cite{ahmed}
and the relevant $\mathcal{PT}$-symmetric transition threshold was
identified as an inequality associating the real and the imaginary
part prefactors. The pertinent inequality here assumes the form
$(b_I-1)^2 \geq 0$ and is generically satisfied (i.e., $\forall ~b_I$),
as can be expected
by the super-symmetric partnership of the potential with a Hermitian
one bearing real eigenvalues for all $b_I$.

As a side remark, we observe
that $V^{(+)}$ is invariant under $b_{I} \Leftrightarrow m$. Further, both
$m$ and $b_I$ are arbitrary real numbers and not integers. Interestingly,
as shown in \cite{bagchi5}, in case $b_I-m$ is not an integer, then
remarkably, the eigenvalue spectrum has two branches:

\begin{equation}\label{3}
    E_n^{(1)} = -(m-n-1/2)^2\,,~~n = 0,1,2,...,n_{max}
\end{equation}
where $m-3/2 \le n_{max}  < m-1/2$, and
\begin{equation}\label{4}
    E_n^{(2)} = -(b_I-n-1/2)^2\,,~~n = 0,1,2,...,n_{max}
\end{equation}
where $b_I -3/2 \le n_{max} < b_l -1/2$.

From this, we infer that when $m=1$ and $b_I$ is not an integer, $H^{+}$ has two nodeless states (i.e. $n=0$) with energy eigenvalues and eigenfunctions

\begin{equation}\label{5}
E_0^{(1)} = -1/4\,,~~\psi_0^{(1)} = \sqrt{\sech(x)} ~e^{2ib_I \tan^{-1}(\tanh x/2)}\,,
\end{equation}
\begin{equation}\label{6}
E_0^{(2)} = -(b_I -1/2)^2\,,~~\psi_0^{(2)} = \sech^{b_I -1/2}(x) ~e^{2i \tan^{-1}(\tanh x/2)}\, ,
\end{equation}
although the latter (as per our spectral results of Fig.~\ref{fig:profile})
will only be present for $b_I>1/2$. Interestingly, while for $1/2 < b_l < 1$, $E_0 = -1/4$ is the ground state, but for $1 < b_l < 3/2$, $E_0 = -(b_l -1/2)^2$ corresponds to the ground state.

\section{Nonlinear Generalization of the Model}

We now turn to the corresponding nonlinear model which is the basis for
the present analysis. Examining the case of the focusing nonlinearity,
the operator $H^{(+)}$ is augmented into the nonlinear problem:
\begin{eqnarray}
i u_t = H^{(+)} u - |u|^{2\kappa} u .
\label{eq10}
\end{eqnarray}
The most physically relevant case is that of the cubic
nonlinearity $\kappa=1$, corresponding to the Kerr effect, although in
recent years, examples of higher order nonlinearities (like
$\kappa=2$ and $\kappa=3$) have been
experimentally realized; see, e.g., for a recent example~\cite{pernamb}.
The relevant nonlinear problem has been partially
considered for $\kappa=1$ in a two-parametric generalization of the potential
associated with $V^{(+)}$~\cite{ziad2}; see also
the more recent discussions of~\cite{nazari,chen}.
While all of these works were restricted to the cubic case,
below we will obtain exact solutions for arbitrary nonlinearity
powers. Moreover, we will explain, through our $\mathcal{PT}$-SUSY
framework the existence of nonlinear dipole (and, by extension,
tripole etc.) solutions identified in~\cite{chen},
emerging from the higher excited states of the underlying linear problem.
It can be directly found that the relevant
nonlinear single-hump solution for arbitrary $k$ is of the form:
\begin{eqnarray}
u(x,t)=e^{-iEt} A {\rm sech}^{1/\kappa}(x) e^{i \phi(x)} ,
\label{eqn11}
\end{eqnarray}
where
\begin{equation}
    A^{2\kappa}=\frac{4\kappa^2}{(\kappa+2)^2}\left[\frac{(\kappa+2)^2}{4\kappa^2}-m^2\right]\left[\frac{(\kappa+2)^2}{4\kappa^2}-b_I^2\right],
\end{equation}
\begin{equation}
    \phi(x)=\frac{4m\kappa b_I}{\kappa+2} \tan^{-1}\left(\tanh(\frac{x}{2})\right) .
\end{equation}
and
\begin{equation}\label{eq:Ekappa}
    E=-\frac{1}{\kappa^2} .
\end{equation}
Note when $m=0$ and $b_I =1/2$, ~~ $V^{(+)} \rightarrow 0$ and our solution reduces to the well known solution of the NLSE with
$   A^{2\kappa} = (1+ \kappa)/\kappa^2$.  Also note that when $\kappa=2$,
\begin{equation}
A^{2\kappa} \rightarrow (b_I^2-1) (m^2-1),
\end{equation}
which vanishes at either $b_I =1$ or the special case $m=1$.
Hereafter, we again restrain consideration to the special case of
$\kappa=1$.

\begin{figure}
\includegraphics[width=6cm]{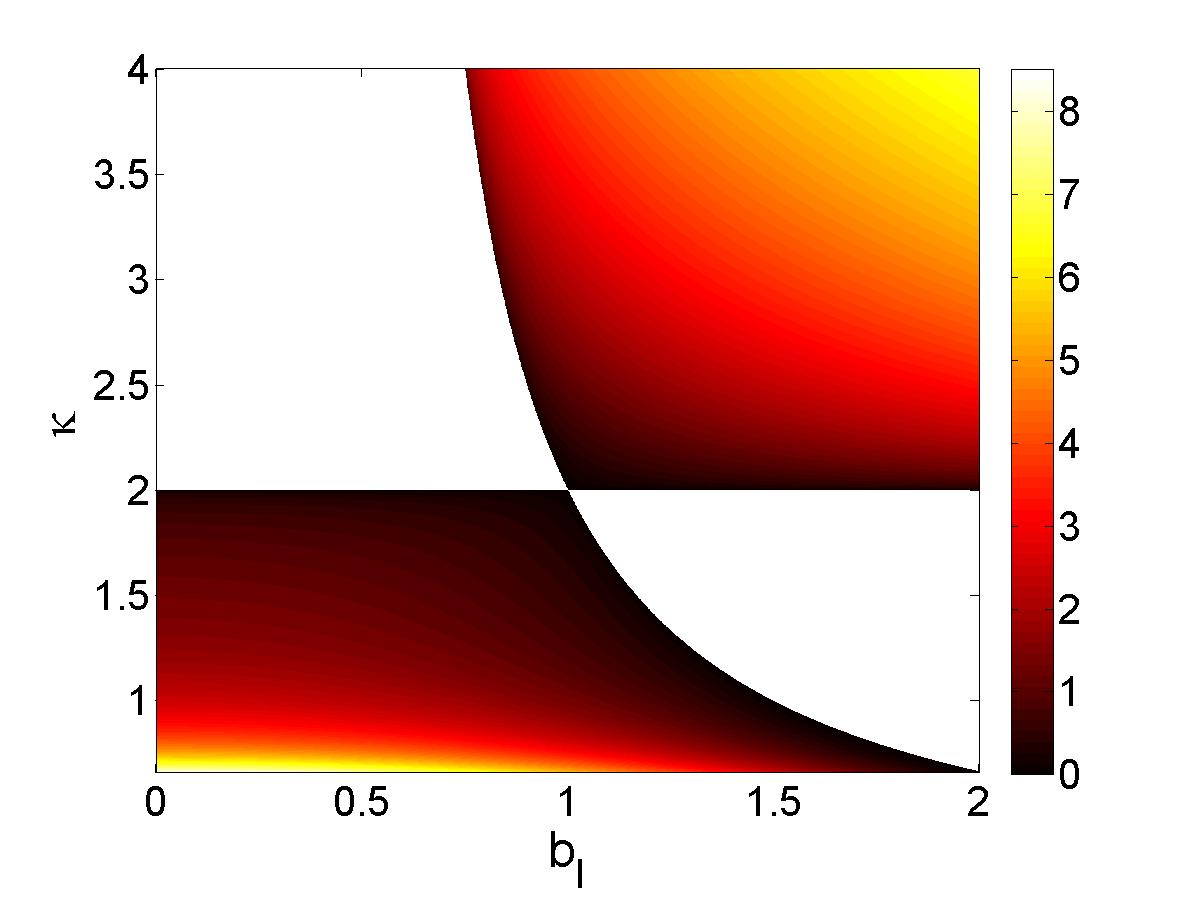}
\caption{Norm of the solutions Eq.~(\ref{eqn11}) as a function of $b_I$ and $\kappa$ when $m=1$. }
\label{fig:norm}
\end{figure}

For $b_I\rightarrow b_{I,c}$, with $b_{I,c}^2=\frac{(\kappa+2)^2}{4\kappa^2}$, the amplitude $A$ tends to zero and the solution (\ref{eqn11}) becomes the solution of the corresponding linear limit (\ref{6}) by virtue of condition (\ref{eq:Ekappa}). The solution (\ref{eqn11}) exists for $b_I<b_{I,c}$ when $\kappa<2$ and for $b_I>b_{I,c}$ if $\kappa>2$. Our analytical
expression only yields the trivial solution  for $\kappa=2$ as mentioned
earlier.
Notice also that when $b_I$ is fixed and $\kappa$ varied, the solution tends to (\ref{5}) when approaching the $\kappa=2$ limit. In addition, Fig. \ref{fig:norm} shows the dependence of norm with respect to $b_I$ and $\kappa$ when the condition (\ref{eq:Ekappa}) for solution (\ref{eqn11}) is applied. The value of the norm is
\begin{equation}
    N=\int_{-\infty}^\infty|u(x,t)|^2\mathrm{d}x=A^2B\left(\frac{1}{2},\frac{1}{\kappa}\right) ,
\end{equation}
where it has been taken into account that $A\in\mathbb{R}$ and $B(x,y)$ is the Euler's beta function.

\begin{figure}
\begin{tabular}{cc}
\multicolumn{2}{c}{\includegraphics[width=6cm]{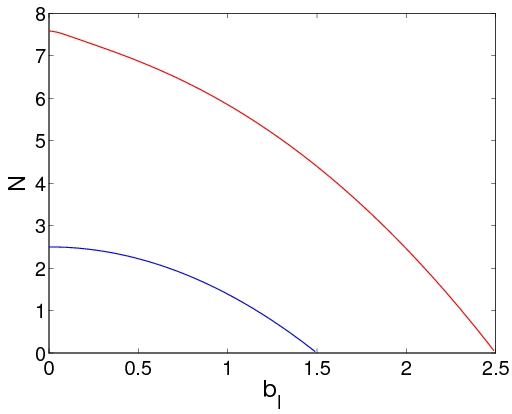}} \\
\includegraphics[width=6cm]{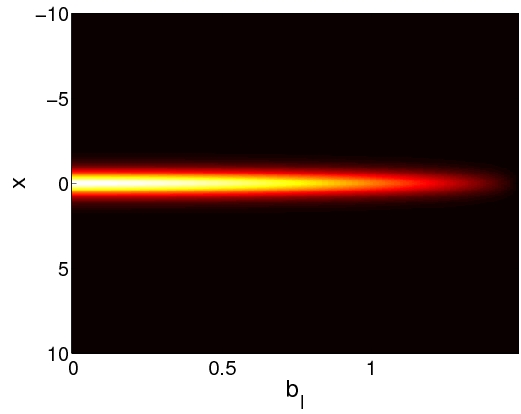} &
\includegraphics[width=6cm]{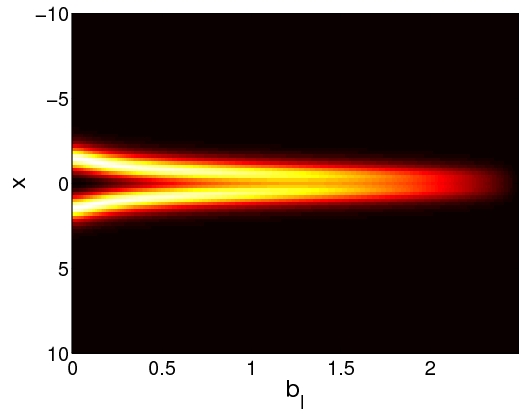} \\
\end{tabular}
\caption{(Top panel) Norm of the solutions with $\kappa=1$ as a function of $b_I$. The blue line corresponds to the nodeless solution whereas the red line corresponds to the ``dipole'' branch (cf.~\cite{chen}) possesing a node.
The bottom left (right) panel showcases the modulus of the nodeless
solution (solution with a node) as a function of $x$ for different values of
$b_I$. It can clearly be seen that the amplitude of the solution goes to
$0$ as $b_I \rightarrow 3/2$ ($b_I \rightarrow 5/2$).}
\label{fig:kappa1}
\end{figure}

\begin{figure}
\begin{tabular}{cc}
\multicolumn{2}{c}{\includegraphics[width=6cm]{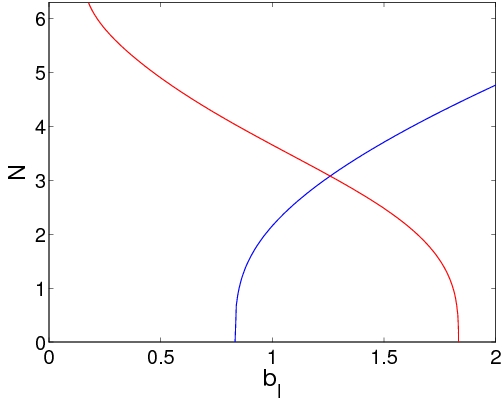}} \\
\includegraphics[width=6cm]{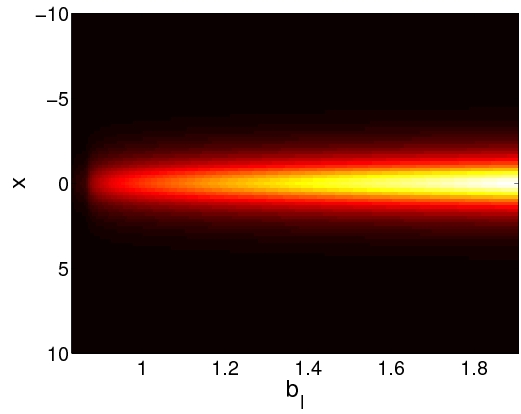} &
\includegraphics[width=6cm]{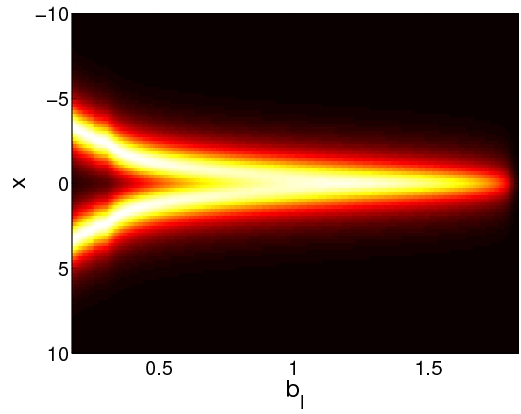} \\
\end{tabular}
\caption{(Top panel) Norm of the solutions with $\kappa=3$ as a function of $b_I$. The blue line corresponds to the nodeless solution whereas the red line
corresponds to the solution with a node (i.e., the ``dipole'').
The bottom left (right) panel
showcases the modulus of the nodeless solution (solution with a node, although
the node itself disappears as $b_I$ increases) as a function of $x$ for different values of $b_I$. It can clearly be seen that the amplitude of the solution goes to $0$ as $b_I \rightarrow 5/6$ ($b_I \rightarrow 11/6$), i.e., the
corresponding linear limit value for $E=-1/9$.}
\label{fig:kappa3}
\end{figure}

Horizontal ``cuts'' along the graph of Fig.~\ref{fig:norm}
are shown in Figs. \ref{fig:kappa1} and \ref{fig:kappa3} where the continuum tendency to the linear limit
(dark) is shown as a variation over $b_I$
for $\kappa=1$ and $\kappa=3$, respectively.
Apart from the analytical solution (\ref{eqn11}) which collides with the nodeless solution of the linear Schr\"odinger equation, we have been able to find numerically the branch that collides with the solution with a node [$n=1$ in (\ref{3})-(\ref{4})]. These solutions are the generalizations (for arbitrary $\kappa$)
of the ``dipoles'' of~\cite{chen}.
In those cases, solutions exist as long as $b_I<b_{I,c}+1$ and
their monotonicity for $\kappa > 2$ is opposite to that of
the fundamental solutions analytically identified above (hence, the above
mentioned collision).
Interestingly,
it is worth mentioning that the latter dipole
branch is present even for $\kappa=2$. In the same spirit,
higher order generalizations (e.g. tripoles, quadrupoles, etc.)
can also be expected in the spirit of~\cite{chen},
degenerating to the linear limit, respectively for
$b_I \rightarrow b_{I,c} +2$, $b_I \rightarrow b_{I,c} + 3$, etc.

\begin{figure}
\begin{tabular}{cc}
\includegraphics[width=6cm]{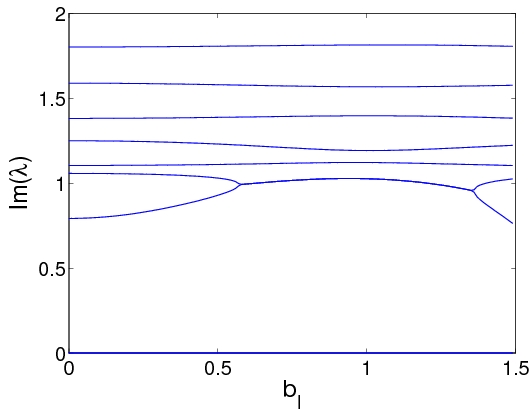} &
\includegraphics[width=6cm]{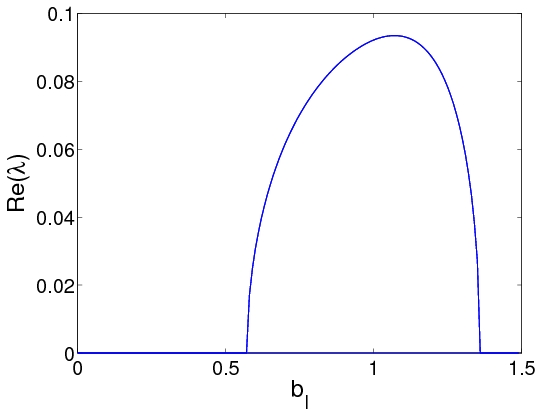} \\
\includegraphics[width=6cm]{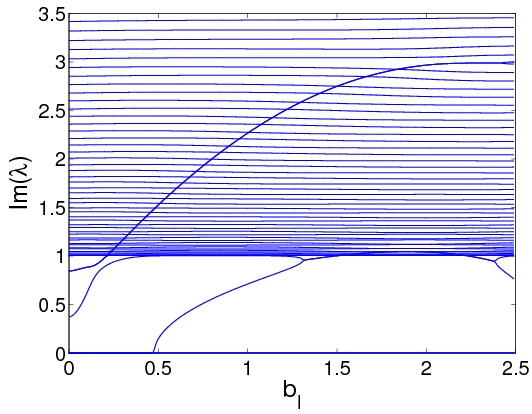} &
\includegraphics[width=6cm]{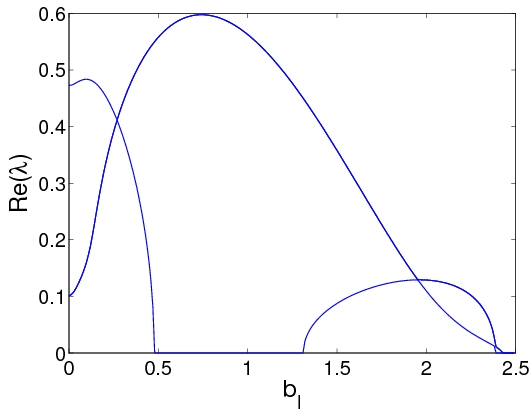} \\
\end{tabular}
\caption{Imaginary part (left panels) and real part (right panels) of the eigenvalues associated with the linearization around the solution of the nodeless (top panels) and the single-node (i.e., dipole) solutions branches (right panel) for $\kappa=1$. It can be observed that the nodeless solutions become unstable in the interval $0.56 < b_I < 1.37$, whereas the solutions with a single node are unstable for all $b_I$ except for a very small interval $2.43<b_I<2.5$ close to the upper existence limit; in addition, for $b_I<0.48$ the
latter waveform is also exponentially unstable.}
\label{fig:kappa1stab}
\end{figure}

We now turn to the detailed stability analysis of the
relevant soliton solutions (which was not explored systematically
in~\cite{ziad2,nazari}, but was touched upon in~\cite{chen} for
$\kappa=1$). In fact, in~\cite{ziad2} a particular case example of an evolution
(cf. Fig. 2 therein), as well as the positivity of the Poynting
vector flux led those authors to conclude that the relevant
solutions were nonlinearly stable. However, our systematic
spectral stability analysis, illustrated in Figs. \ref{fig:kappa1stab} and \ref{fig:kappa3stab}, indicates otherwise.
In particular, we use a linearization ansatz of the form
\begin{eqnarray}
u(x,t)=e^{iEt} \left[u_0(x) + \left(a(x) e^{\lambda t}
+ b^{\star}(x) e^{\lambda^{\star}t} \right) \right] ,
\label{eqn12}
\end{eqnarray}
where $u_0(x)$ is the spatial profile of the standing wave
solution of Eq.~(\ref{eqn11}), while $\{a(x),b(x)\}$ and
$\lambda$ correspond, respectively, to the eigenvector
and eigenvalue of the linearization around the solution.
The existence of eigenvalues with Re$(\lambda) > 0$
would in this ($\mathcal{PT}$-symmetric and hence still ensuring
the quartet symmetry of the relevant eigenvalues) context
signal the presence of an instability.

\begin{figure}
\begin{tabular}{cc}
\includegraphics[width=6cm]{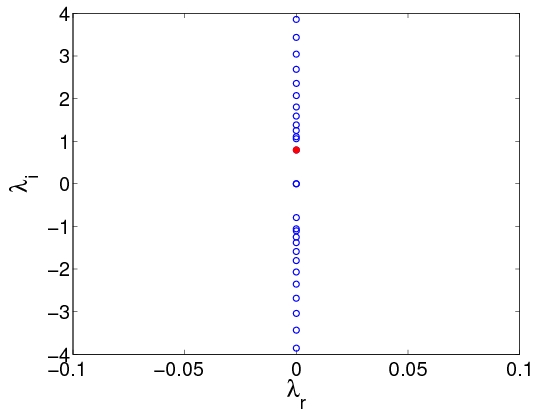} &
\includegraphics[width=6cm]{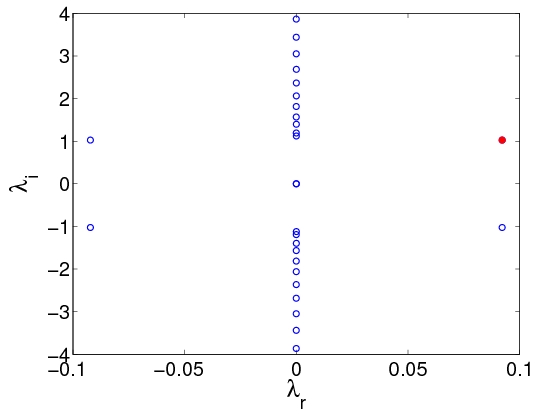} \\
\end{tabular}
\caption{Two case examples of the spectral plane $(\lambda_r,\lambda_i)$ of eigenvalues $\lambda=\lambda_r + i \lambda_i$ of the solution for $b_I=0$ (left panel) and $b_I=1$ (right panel). The eigenvalue which is spectrally stable in the left panel but whose collision with the band edge of the continuous spectrum is responsible for the instability in the right panel is denoted by a red mark.}
\label{fig:profile3}
\end{figure}

\begin{figure}
\begin{tabular}{cc}
\includegraphics[width=6cm]{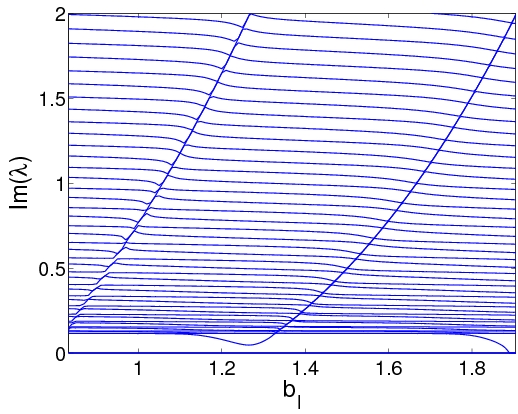} &
\includegraphics[width=6cm]{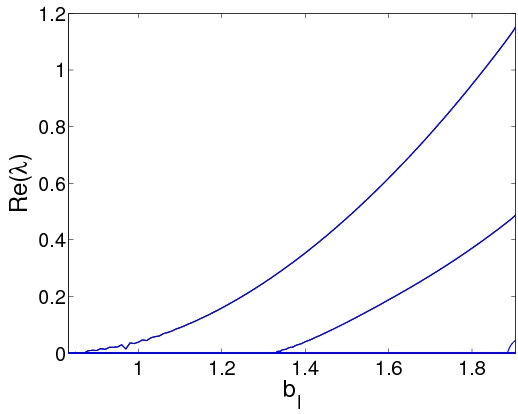} \\
\includegraphics[width=6cm]{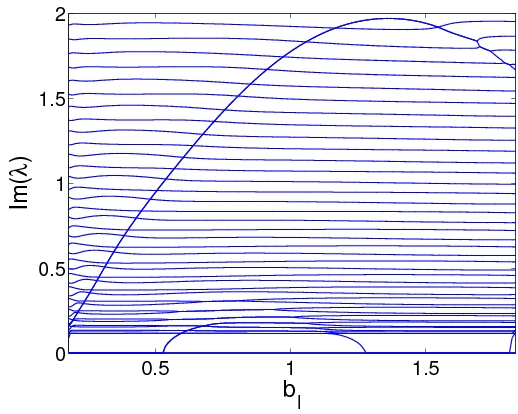} &
\includegraphics[width=6cm]{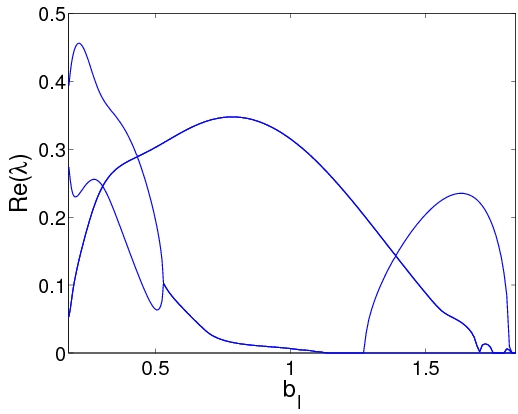} \\
\end{tabular}
\caption{Imaginary part (left panels) and real part (right panels) of the eigenvalues associated with the linearization around the solution of the nodeless (top panels) and the single-node solution branches (right panel) for $\kappa=3$. It can be observed that all the solutions (for different values
of $b_I$) are unstable; see the text for a detailed description of the
eigenvalue variation over $b_I$.}
\label{fig:kappa3stab}
\end{figure}

We can see in Fig.~\ref{fig:kappa1stab} that indeed such an
instability is present in the interval $0.56 < b_I < 1.37$
for the nodeless solutions of $\kappa=1$.
Further examination of the relevant phenomenology in Fig.~\ref{fig:profile3}
reveals the origin of the instability and its {\it stark} contrast
with the corresponding phenomenology in the standard NLS model.
In particular, the breaking of translational invariance
(due to the presence of the potential) leads the corresponding
neutral mode to exit along the imaginary axis of the spectral
plane $(\lambda_r,\lambda_i)$ of the eigenvalues $\lambda= \lambda_r +
i \lambda_i$. However, it is well-known~\cite{kks} that in the standard
Hamiltonian case, the relevant ``internal mode'' of the solitary wave
associated with translation has a positive energy or
signature and hence its collisions with other modes, including
ones of the continuous spectrum, do not lead to instability.
Here, however, as illustrated in Fig.~\ref{fig:profile3} exactly
the opposite occurs. As the parameter $b_I$ is varied, the relevant
eigenvalue moves towards the continuous spectrum (whose lower
limit is $\lambda= \pm i$) and the collision with it leads to
a complex eigenvalue quartet, a feature that would never be possible
for a single soliton of the regular NLS, under a translation-symmetry-breaking
perturbation. This is a remarkable feature of the $\mathcal{PT}$-symmetric
NLS model that is worthy of further exploration, possibly utilizing the
notion (recently discussed for $\mathcal{PT}$-symmetric models in~\cite{nixon})
of Krein signature. Notice that the work of~\cite{nixon}
considered a case in the vicinity of the $\mathcal{PT}$-phase transition,
whereas in our setting, such a phase transition is {\it impossible},
given the real nature of the super-partner P{\"o}schl-Teller
potential, as discussed above.

Fig. \ref{fig:kappa1stab} shows that dipolar
solutions with a single node are unstable for almost all of their
existence interval expect when $2.43<b_I<2.5$, i.e. in the immediate
vicinity of the linear limit. There are three different instability intervals: (1) for $b_I\leq0.48$ there are two instabilities, one of exponential nature and an oscillatory one, for $0.48<b_I\leq1.31$ the oscillatory instability is the only one that persists, while the formerly real eigenmode crosses the
spectral plane origin
and becomes imaginary for larger $b_I$.
(3) for $1.31<b_I<2.43$, there are two oscillatory instabilities, the previously mentioned one, and another one caused (in a way similar to
the nodeless case) by the climbing up the imaginary axis of the eigenmode
formerly unstable as a real pair, and its eventual collision with an
eigenvalue bifurcating from the continuous spectrum.

For nodeless solutions with $\kappa=3$, we can observe in
Fig.~\ref{fig:kappa3stab} (top panels)
that the solution is unstable
throughout its range of existence because of an eigenmode
entering the phonon band and causing oscillatory instabilities; a second localized mode enters at $b_I=1.33$ and, finally, for $b_I=1.89$, the soliton becomes exponentially unstable.
Analogously, the solutions with a node are unstable for all $b_I$, incurring,
in fact, typically multiple instabilities for each value of the
parameter, which can be summarized as follows (see the bottom panels of Fig.~\ref{fig:kappa3stab}): an oscillatory instability is present for almost every value of $b_I$ ($b_I\lesssim1.82$);
apart from this we observe, for low values of $b_I$, two pairs of real eigenvalues which coalesce into a quartet at $b_I\approx0.525$; this quartet
leads to two imaginary pairs when  $b_I\approx1.14$; one of
them moves down along the imaginary axis and finally, at $b_I
\approx 1.27$, an additional instability due to a real pair emerges..

It is relevant to note in passing, another interesting result which
relates to the $\kappa<1$ case; we have found that for $\kappa<2/3$, the nodeless soliton is stable for {\it every} $b_I$. This, as well as the results
above indicate the strong dependence of the stability properties on the
precise strength of the nonlinearity parameter.

\begin{figure}
\begin{tabular}{cc}
\includegraphics[width=6cm]{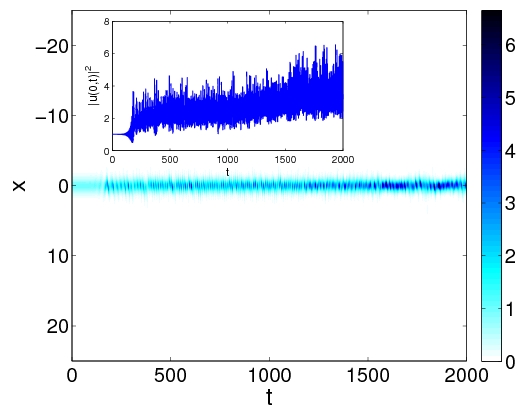} &
\includegraphics[width=6cm]{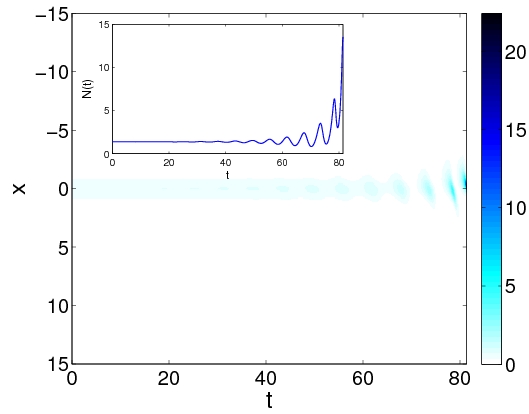} \\
\includegraphics[width=6cm]{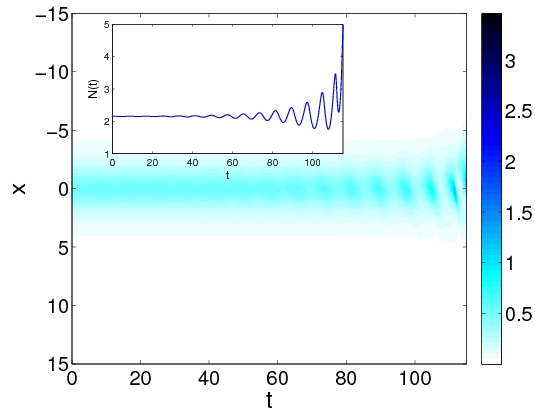} &
\includegraphics[width=6cm]{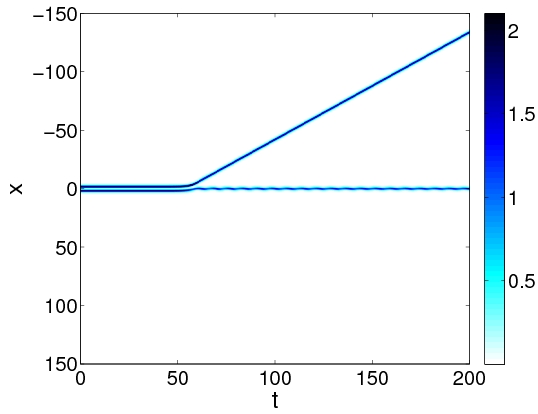} \\
\end{tabular}
\caption{(Top left panel) Space-time contour plot of the evolution of the squared modulus (density) of the solution during its unstable dynamics for $b_I=0.65$ and $\kappa=1$; the inset shows the evolution of the (maximal) density of the solution occurring at $x=0$. The top right (bottom left) panel shows the
evolution of the density for $b_I=1$ and $\kappa=1$ ($\kappa=3$); the inset
shows the evolution of the norm and displays its eventual indefinite growth.
The bottom right panel considers the evolution of the unstable soliton with a
node for $b_I=0.2$ and $\kappa=1$, leading eventually to a split of
the two humps into a stationary (at $x=0$) and a traveling one
(at $x<0$).}
\label{fig:simul}
\end{figure}

Finally, we consider the dynamics of these unstable waveforms for several prototypical cases in Fig.~\ref{fig:simul}. For $b_I=0.65$ and $\kappa=1$ we observe that when $t\gtrsim200$, the oscillatory (as predicted by our eigenvalue computations) nature of the instability gradually kicks in and eventually renders the solitary wave more highly localized (i.e., narrower) at $x=0$ and with a larger amplitude (i.e., taller). Subsequently, the amplitude of the pulse is subject to breathing, but it remains fairly robust, even after multiple collisions with small amplitude radiative wavepackets scattering back and forth from the boundaries
(not visible in the scale of the plot).
For $b_I=1$ and $\kappa=1$, the growth rate is larger and the instability effects stronger; it manifests in an oscillatory growth of the soliton (given
the oscillatory nature of the instability), as well as a ``swinging''
of the solution between the gain ($x<0$) and loss ($x>0$) regions,
according to the terminology of~\cite{nazari}. This eventually
leads to rapid growth (beyond the resolution of the numerical
scheme). We do not follow the solution past these large values
of its amplitude.
This behaviour is generic for the oscillatory instabilities as long as the growth rate is above a threshold, as shown also in the example for $b_I=1$ and $\kappa=3$, and for the nodeless and single-node solitons.
Finally, we have considered the effect of the exponential instabilities
in solitons with a node and $\kappa=1$. Those solitons are both exponentially and oscilatorily unstable for $b_I<0.48$. In that interval, the soliton is double-humped (see Fig.~\ref{fig:kappa1}). In the particular example of Fig.~\ref{fig:simul}, we have taken $b_I=0.2$ where the exponential instability dominates to the oscillatory one. The dynamics here can be described as follows:
the hump originally located at the loss ($x>0$) side shifts to and remains pinned with regular oscillations of the amplitude at $x=0$.
On the other hand,  the hump initially at the gain ($x<0$) side
is ``ejected'', as  a result of the instability, along the
(exponentially localized around $x=0$) gain side of $x<0$.

\section{Conclusions \& Future Challenges}

In the present work, we revisited a potential that
has been explored previously in a number of studies
relating to $\mathcal{PT}$-symmetric models. We discussed how
for a special monoparametric family within this model,
it is not only $\mathcal{PT}$-symmetric but also super-symmetric
with a partner which is the P{\"o}schl-Teller potential,
a feature which enabled us to identify its purely
real spectrum
(and the bifurcations of bound states within it) and to
corroborate the corresponding results numerically.
As a byproduct of its super-symmetric origin, this family
of potentials was found to be devoid of a $\mathcal{PT}$-phase-transition.
We then turned to a nonlinear variant of the model for
arbitrary powers of the nonlinearity,
and illustrated that exact nonlinear solitonic solutions
degenerated in the appropriate limit to the linear states
identified previously. While there was no $\mathcal{PT}$-phase-transition
in this model, we found that the nonlinear solutions were
still subject to instabilities, such as e.g. the one
stemming from a collision
of an internal mode with the continuous spectrum (band edge),
leading to a quartet of eigenvalues. The ensuing oscillatory
instability led to an oscillating, progressively larger
amplitude soliton
in the cases examined. Additional families of solutions were discussed,
including e.g. the one-node branch (dipolar solution), and their
reduced stability (in comparison to the nodeless branch) was
illustrated.

While this work, to the best of our understanding, is only
a first step in connecting all three notions of $\mathcal{PT}$-symmetry,
super-symmetric potentials and nonlinear phenomenology (including
instabilities), naturally this is a theme that is worthy of
considerable further studies. For one thing, there are numerous
additional super-symmetric potentials with real spectra that
can be devised and are worth examining. For instance, the
sl$(2,C)$ considerations of~\cite{bagchi} already suggest
some such options including the super-potentials
$W(x)=(m-1/2) {\rm coth}x-i b_I {\rm cosech}(x)$
or $W(x)=\pm (m-1/2)- i b_I \exp({\mp x})$. Additionally, there have
already been proposals for $\mathcal{PT}$-symmetric square well
potentials considered in the SUSY framework~\cite{bagchi3}, and
for non-Hermitian SUSY hydrogen-like Hamiltonians with
real spectra~\cite{bagchi4}. Especially in the latter higher dimensional
context, understanding the delicate interplay of $\mathcal{PT}$-symmetry,
super-symmetric models with their bound states, and collapse
induced by nonlinearity could be an especially interesting
topic. Finally, there are some potentially intriguing deeper
connections. SUSY partners are based on commutation formulae
as are integrable nonlinear equations. Perhaps the latter
is intrinsically responsible for the similarity of the
structure of the SUSY partner potentials with the well-known
Miura transformation responsible for converting the modified
Korteweg-de Vries equation to the Korteweg-de Vries equation~\cite{drazin}.
Exploring these connections further would constitute an important
direction for further studies and efforts along this direction
are already underway~\cite{olshanii}.

\section{Acknowledgments} This work was supported in part by the U.S.
Department of Energy.  AK wishes to thank Indian National Science Academy
(INSA) for the award of INSA Senior Professor position at Pune University.
PGK gratefully acknowledges support from NSF-DMS-1312856,
the Binational Science Foundation under
grant 2010239, and the ERC under FP7, Marie Curie Actions, People,
International Research Staff Exchange Scheme (IRSES-605096).

\end{document}